\documentstyle[emulateapj,psfig,apjfonts]{article}

\def\gs{\mathrel{\raise0.35ex\hbox{$\scriptstyle >$}\kern-0.6em 
\lower0.40ex\hbox{{$\scriptstyle \sim$}}}}
\def\ls{\mathrel{\raise0.35ex\hbox{$\scriptstyle <$}\kern-0.6em 
\lower0.40ex\hbox{{$\scriptstyle \sim$}}}}
\submitted{Received --- ; Accepted ---}

\lefthead{Kodama, T. \& Matsushita, K.}
\righthead{Optical vs. X-ray Properties in Early-Type Galaxies}

\begin{document}

\title{Homogeneity of Stellar Populations in Early-Type Galaxies with
 Different X-ray Properties}

\author{Tadayuki Kodama\altaffilmark{1,2} \& Kyoko Matsushita\altaffilmark{3}
}
\affil{\small 1) Department of Physics, University of Durham, South Road, 
Durham DH1 3LE, UK}
\affil{\small 2) Department of Astronomy, University of Tokyo, Hongo,
Bunkyo-ku, Tokyo 113-0033, Japan}
\affil{\small 3) Department of Physics, Tokyo Metropolitan University,
1-1 Minami-Ohsawa Hachioji, Tokyo 192-0397, Japan}

\begin{abstract}
We have found the stellar populations of early-type galaxies are homogeneous
with no significant difference in color or Mg2 index, despite the dichotomy
between X-ray extended early-type galaxies and X-ray compact ones.
Since the X-ray properties reflect the potential gravitational structure
and hence the process of galaxy formation, the homogeneity of the stellar
populations implies that the formation of stars in early-type galaxies predates
the epoch when the dichotomy of the potential structure was established.
\end{abstract}


\section{Introduction}

ASCA X-ray observations of NGC 4636 (Matsushita et al. 1998) and some
other giant early-type galaxies (Matsushita 1997) show that
early-type galaxies can be classified into two categories in terms of
X-ray extent.
Some early-type galaxies
have a very extended dark matter halo characterized by X-ray
emission out to $\sim$ 100~kpc from the galaxy center, while others have a
compact X-ray halo. The galaxies with an extended X-ray emission
can be interpreted as sitting in larger scale potential structure,
such as galaxy groups, subclumps of clusters, or clusters themselves,
as well as sitting their own potential well associated with each galaxy.

Potential structure must have played a big role in the cource of galaxy
formation.
If the difference in potential structure had been already established
before the bulk of stars formed, we would expect some differences in stellar
populations, such as mean age or metallicity, as well.
A deeper potential well would keep the gas more effectively against the
thermal energy input by supernova (SN) explosion,
and the chemically enriched gas can be recycled more efficiently,
and the galaxy would end up with higher mean stellar metallicity (Larson 1974).
Therefore we would expect that the X-ray extended galaxies
have higher metallicities than the X-ray compact ones at a given stellar mass.
Furthermore, considering that the higher density peaks collapse earlier in the
Universe, which is likely to be the case for the
X-ray extended galaxies sitting in the local density peaks,
we would also expect their older ages than those of the X-ray compact ones.
Both of these effects would make the colors of the X-ray extended galaxies
redder.
The central question of this paper is, therefore, how this dichotomy in X-ray
properties hence the potential structure is related to the optical properties
which trace the stellar populations in galaxies.

Another interesting issue is whether the number of globular clusters per
unit optical luminosity of galaxy correlates with the X-ray extent of galaxy.
This is because, if the X-ray extended galaxies are the products of galaxy
mergers as they are located in the center of larger scale potential structure,
and if a considerable number of new globular clusters form during galaxy
mergers as suggested by Zepf and Ashman (1993),
we would expect more globular clusters in the X-ray extended galaxies
for a given optical luminosity.

Matsushita (2000; hereafter M2000) has recently compiled the X-ray
properties of 52 nearby early-type galaxies from ROSAT data.
Together with the archival data of various optical properties,
we now compare the optical properties with the X-ray properties
to examine the correlation between them.

The structure of this paper is the following.
In \S~2 we summarize the X-ray properties of our sample of early-type galaxies,
highlighting the dichotomy of the potential structure.
In \S~3 we present their optical properties, including integrated colors and
Mg$_2$ index. We show the homogeneity
of the stellar populations despite the dichotomy in X-ray properties.
We discuss the impact of this result on the formation of early-type galaxies
in \S~4, and conclude the paper in \S~5.
We use $H_0=75$ km s$^{-1}$ Mpc$^{-1}$ throughout this paper.

\section{X-ray Properties}

We use the same sample of early-type galaxies presented in M2000.
The sample is composed of 52 bright early-type galaxies, of which 42 are
ellipticals and 10 are S0 galaxies.
M2000 have selected all the early-types observed by PSPC, available
in the ROSAT archival data,
whose $B$-band magnitudes are brighter than 11.7.
The environment of the sampled galaxies varies from cluster environment
(Virgo, Fornax, and Centaurus clusters) to galaxy groups and the field.

Figure~1 shows X-ray luminosity of the inter stellar medium (ISM) within
a radius of 4$r_e$ ($L_X$) against $L_B\sigma^2$ for all the sample
galaxies, where $r_e$, $L_B$ and $\sigma$ indicate 
the effective radius in the optical profile,
the galaxy luminosity in $B$-band (taken from Tully 1988),
and the central velocity dispersion of stars, respectively.
In order to exclude the contribution from low mass X-ray binaries and the
active galactic nuclei, the ROSAT PSPC spectrum (0.2-2.0~keV) is fitted
with two components; soft ($\sim$1~keV) and hard (10~keV),
and only the soft component is used to determine the ISM X-ray luminosity
(M2000).
The quantity $L_B\sigma^2$ is proportional to the kinetic energy of the gas
supplied from stellar mass loss and heated up by random stellar motions.
The solid line, $\log L_X/(L_B\sigma^2)$$=$25.15 (const),
corresponds to the energy balance between the cooling by X-ray emission and
the heating by stellar mass loss, assuming a relation between mass
loss rate and $L_B$ (Ciotti et al. 1991; M2000).
There is a considerable scatter in $L_X$ for a given $L_B\sigma^2$.

\hbox{~}
\centerline{\psfig{file=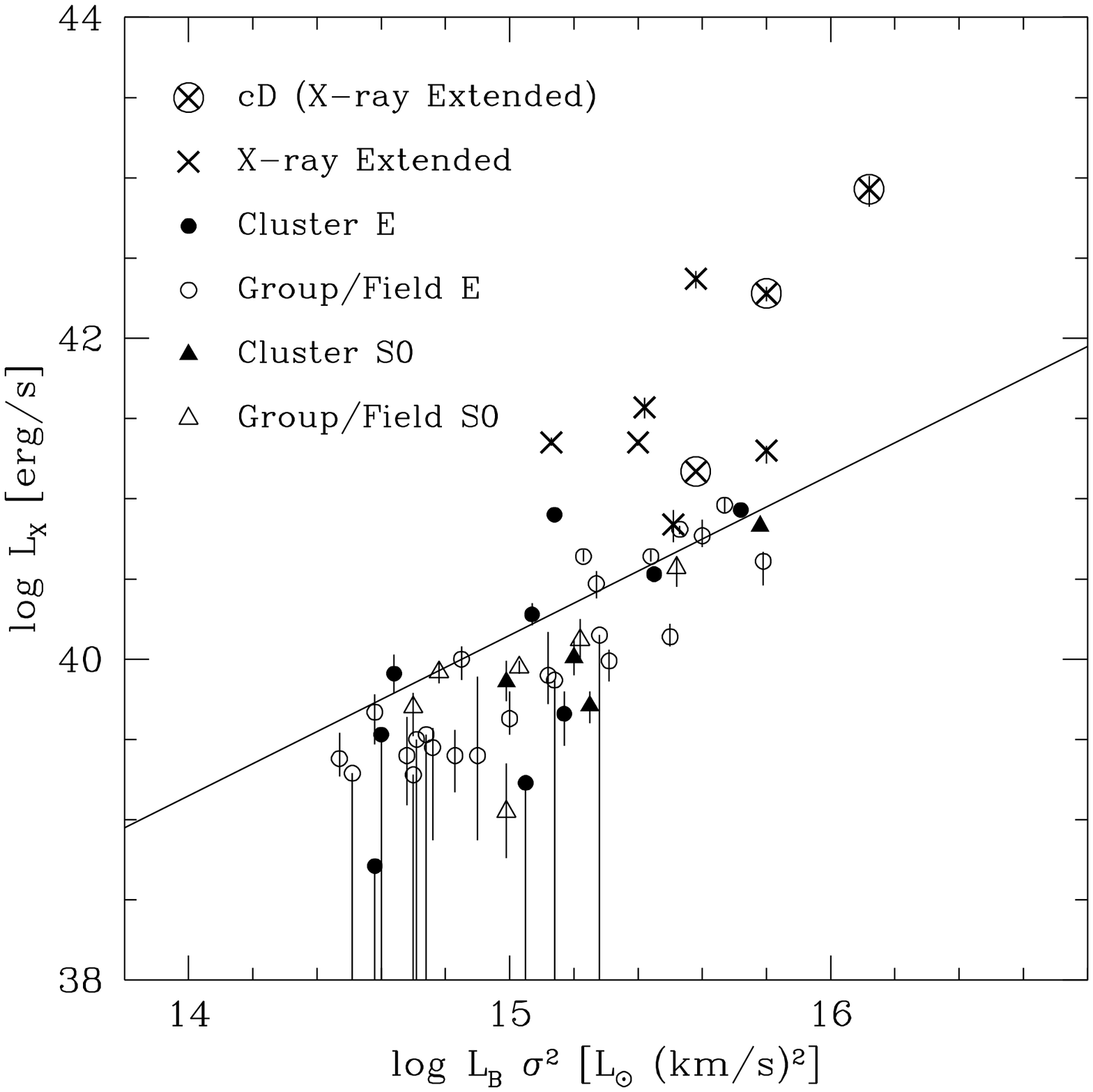,width=2.8in,angle=0}}

\vskip 2mm

\noindent{
\small \addtolength{\baselineskip}{-2pt} 
Fig.~1.\
X-ray luminosity ($L_X$) within 4~$r_e$ versus $L_B\sigma^2$.
Vertical error-bars show two sigma errors in $L_X$.
Crosses indicate the X-ray extended galaxies, and those surrounded by
big circles are the cD galaxies.
The galaxies in cluster environment are shown by filled symbols,
while those in the field or in small groups are shown by open symbols.
Ellipticals are indicated by circles, while S0's are indicated by triangles.
The solid line corresponds to $\log L_X/(L_B\sigma^2)=25.15$ (constant),
on which the X-ray luminosity is just comparable to the kinetic energy
input to the ISM by stellar mass loss and the heating of the ejected gas
by random stellar motions.

\vskip 5mm
\addtolength{\baselineskip}{2pt}
}

\noindent
Many galaxies follow the solid line,
suggesting that their X-ray luminosities can be simply explained by the energy
input from stars through mass loss.
However some galaxies have significantly higher $L_X/L_B\sigma^2$ ratios,
requiring an excess energy to explain such high X-ray luminosities.
Many of these $L_X$ bright galaxies are classified as X-ray extended
galaxies in M2000 (crosses) as they show spatially extended X-ray emission.
They are likely to have more extended dark matter halos residing
on larger scale structure, such as a group or a cluster of galaxies,
as well as on its own galaxy.
Furthermore, the X-ray extended galaxies show significantly higher ISM
temperatures at $r>r_e$ than that of the X-ray compact ones
as a result of their extended potential.
In fact, their mean ISM temperature within 4$r_e$ is about factor of 2
higher than the others at a given $\sigma$ (M2000).
We reproduce this plot in Fig.~2.
We refer these differences as a dichotomy of X-ray properties.
Therefore we will use a quantity $E_X$=$\log L_X/(L_B\sigma^2)$$-$25.15,
the excess energy in $L_X$, as a measure of the `X-ray extent'
later in \S~3. Larger $E_X$ means that a galaxy has a deeper and more extended
potential.
The galaxies which are classified as X-ray extended ones in
M2000 are the following nine:
NGC~4406 (bright galaxy in Virgo cluster),
NGC~4472 (brightest galaxy in Virgo), 
NGC~4486 (cD in Virgo),
NGC~4636 (bright galaxy in Virgo),
NGC~4696 (cD in Centaurus cluster), 
NGC~1399 (cD in Fornax cluster),
NGC~1407 (group center),
NGC~5044 (group center),
and NGC~5846 (group center).

\section{Optical Properties}

We have compiled the optical properties of our sample galaxies
from various sources.
The integrated colors are taken from de Vaucouleurs et al. (1991; RC3).
Following Schweizer \& Seitzer (1992), we defined $(U-V)_{e,0}$ as
\begin{equation}
 (U-V)_{e,0} = (U-V)_{e} + [(U-V)_{T,0}-(U-V)_{T}],
\end{equation}

\hbox{~}
\centerline{\psfig{file=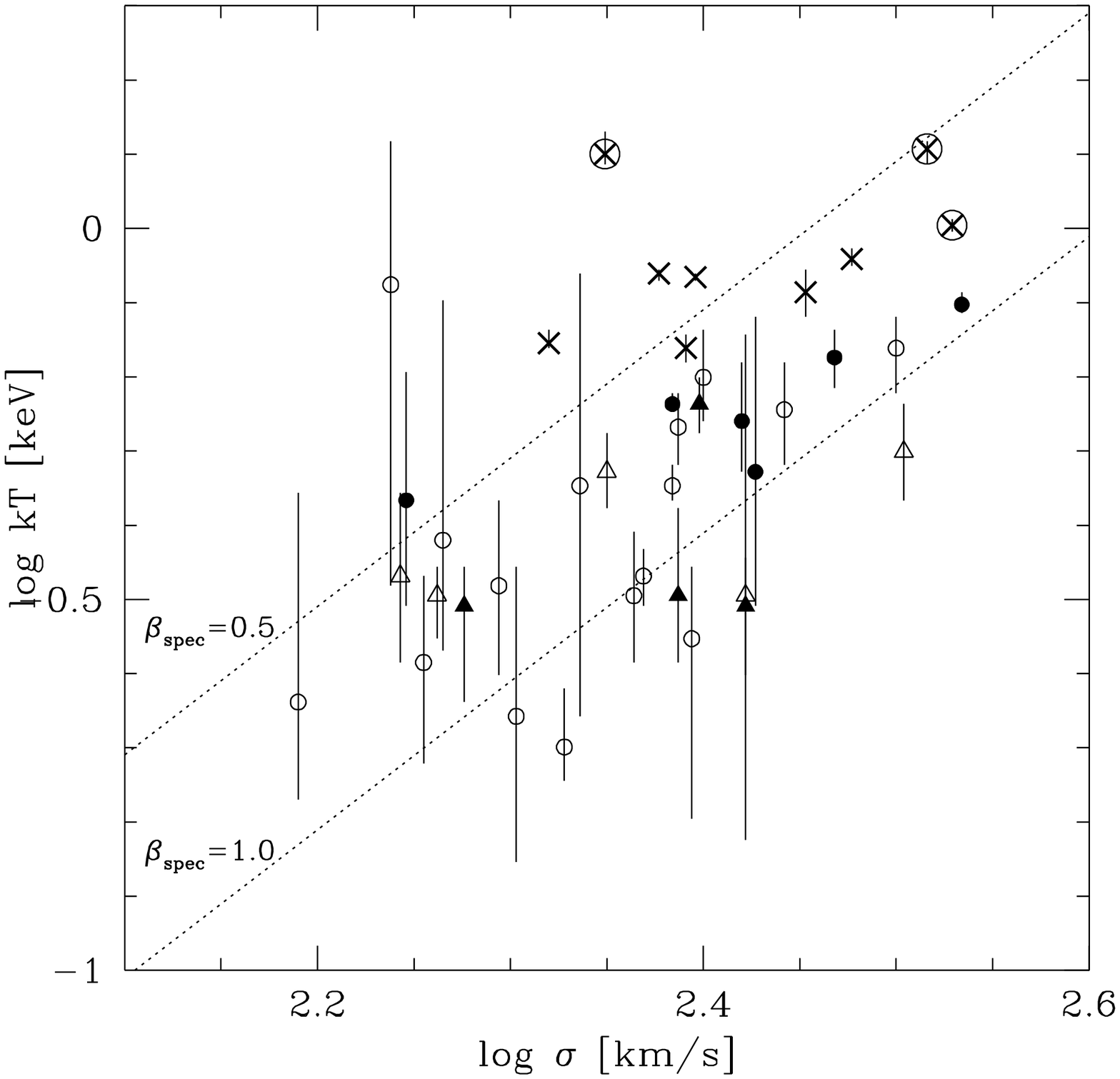,width=2.8in,angle=0}}

\vskip 2mm

\noindent{
\small \addtolength{\baselineskip}{-2pt} 
Fig.~2.\
The ISM temperature ($kT$) within 4~$r_e$ versus central
stellar velocity dispersion ($\sigma$).
Crosses indicate the X-ray extended galaxies, and those surrounded by
big circles are the cD galaxies.
The galaxies in cluster environment are shown by filled symbols,
while those in the field or in small groups are shown by open symbols.
Ellipticals are indicated by circles, while S0's are indicated by triangles.
Two dotted lines correspond to $\beta_{\rm spec}$=0.5 (upper)
and $\beta_{\rm spec}$=1.0 (lower), respectively, where
$\beta_{\rm spec}$ is the ratio between kinetic temperature of stars and
the gas temperature.

\vskip 5mm
\addtolength{\baselineskip}{2pt}
}

\noindent
where subscript $T$ denotes global colors, subscript $0$ indicates
colors corrected for Galactic extinction and redshift, and 
subscript $e$ denotes average colors within the effective radius ($r_e$).
$(B-V)_{e,0}$ is also defined in a similar way.
The integrated colors within $r_e$ in the Cousins system,
$(V-R)_e$ and $(V-I)_e$, are taken from Buta \& Williams (1995).
Galactic extinction is corrected for using the extinction in $B$-band
($A_g$) from RC3 and the extinction curve from Rieke \& Lebofsky (1985).
The reddening corrected colors are denoted as $(V-R)_{e,0}$ and $(V-I)_{e,0}$.
The reason why we use colors within $r_e$ rather than the total colors
is that there are more galaxies available with $r_e$ apertures. This improves
the statistics. Furthermore, the colors within $r_e$ would be more reliable
than the total colors, although no errors are given in the literature for
the total colors.
However, we have checked that the results in this paper did not
change even if we used the total colors.
The central line index Mg$_2$, central velocity dispersion ($\sigma$),
maximum rotation velocity ($V_m$), and deviation from the Fundamental Plane
($\Delta{\rm FP}$) are obtained from Prugniel \& Simien (1996).
A velocity dispersion for NGC~4696 is added from Faber et al. (1989).
The globular cluster specific frequency ($S_N$), which is the number of
globular clusters per unit galaxy luminosity,
is acquired from Ashman \& Zepf (1998).
Finally, the $a_4$ index, which is the fourth cosine coefficient of the Fourier
expansion of isophotal deviation from a pure ellipse, and the ellipticity
($\epsilon$) are taken from Bender et al. (1989).
We will use these values later in this section.
The number of galaxies which have optical properties available are
44 ($U-V$), 47 ($B-V$), 32 ($V-R$), 32 ($V-I$), 43 (Mg$_2$),
49 ($\Delta{\rm FP}$), 44 ($V_m$), 23 ($S_N$), and 25 ($a_4$)
out of a total of 52 galaxies in our X-ray sample.

In the top three panels of Fig.~3, we have plotted
$U-V$, $V-I$ colors and Mg$_2$ index against $\log \sigma$. 
The error-bars indicate one sigma measurement errors.
The error-bars are not shown for Mg$_2$, but are negligibly small ($<$0.005).
The typical error for $\log\sigma$ is $\sim$0.02 (Prugniel \& Simien 1996).
We find the scaling

\hbox{~}
\centerline{\psfig{file=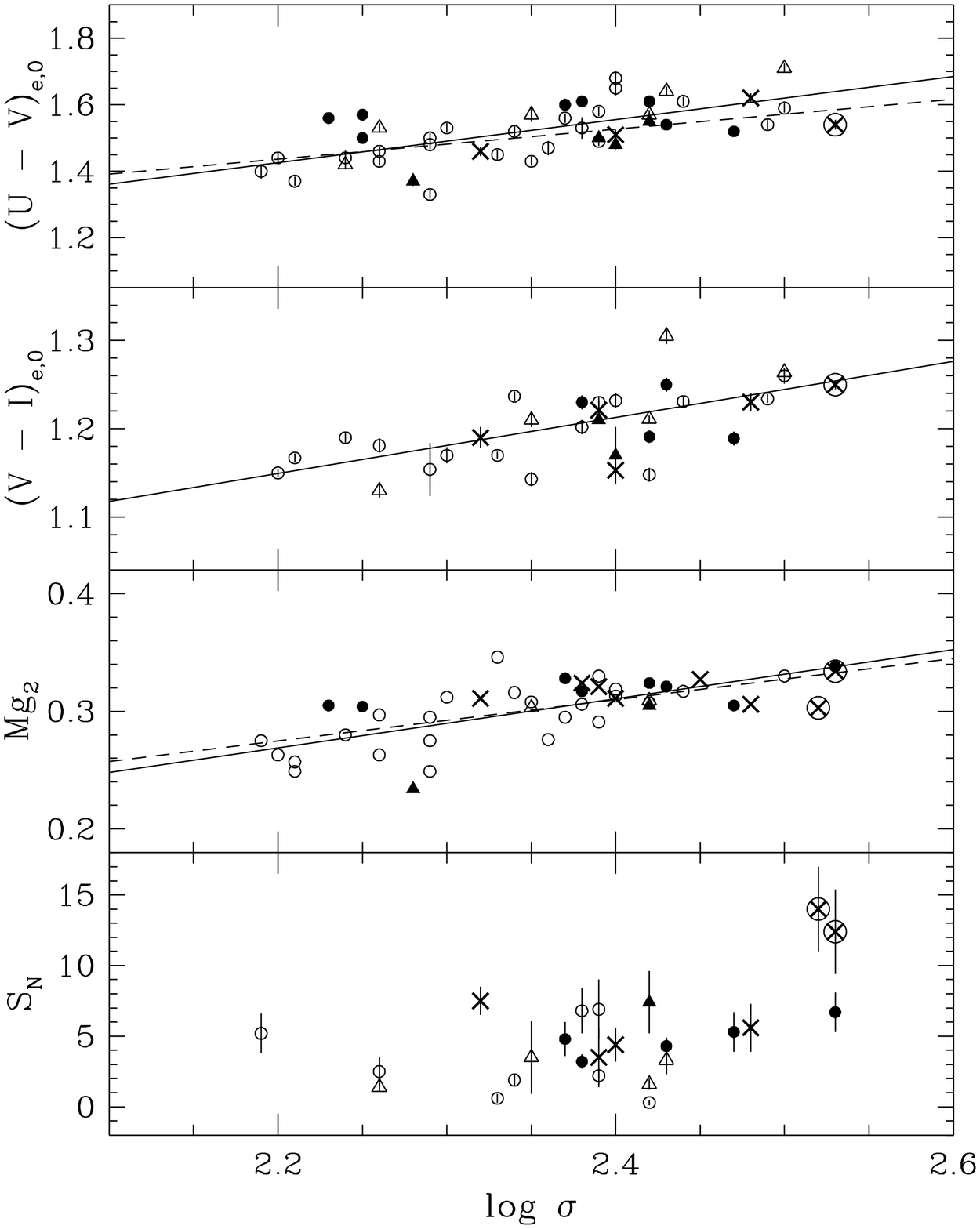,width=3.0in,angle=0}}

\vskip 2mm

\noindent{
\small \addtolength{\baselineskip}{-2pt} 
Fig.~3.\
Integrated colors within an effective radius $(U-V)_{e,0}$ and $(V-I)_{e,0}$,
central Mg$_2$ index, and globular cluster specific frequency $S_N$
(from top to bottom) are
plotted against the central stellar velocity dispersion $\sigma$.
Vertical error-bars (one sigma) are shown except for the Mg$_2$ index.
Crosses indicate the X-ray extended galaxies, and those surrounded by
big circles are the cD galaxies.
The galaxies in cluster environment are shown by filled symbols,
while those in the field or in small groups are shown by open symbols.
Ellipticals are indicated by circles, while S0's are indicated by triangles.
The solid lines are the linear regression lines fitted to the data
excluding the X-ray extended galaxies.
The dashed lines show the relations taken from the literature
(Bower et al. 1992; Burstein et al. 1988).

\vskip 5mm
\addtolength{\baselineskip}{2pt}
}

\noindent
relations for the early-type galaxies as a whole
which are shown by the solid lines.
The linear regression lines are fitted to the data excluding the X-ray
extended galaxies in order to see the difference between the X-ray
extended galaxies (crosses) and the X-ray compact ones (the others),
if present.
For comparison, the same relations taken from Bower, Lucey \& Ellis (1992)
and Burstein et al. (1988) are also reproduced by the dashed lines
on the top panel and the third one, respectively.
Our regression lines agree very well with those in the literature.
Most important, none of the $U-V$, $V-I$, and Mg$_2$ indices of the X-ray
extended galaxies looks different from that of the X-ray compact
ones, being well within the scatter at fixed $\sigma$.

$S_N$ is plotted against $\log \sigma$ in the bottom panel of Fig.~3.
The error-bars correspond to one sigma.
Again, the X-ray extended galaxies do not have systematically different $S_N$
values, except the two circled galaxies, NGC~1399 and NGC~4486
(the former has a larger $\sigma$),
which have a factor of 2-3 as many globular clusters for a given optical
galaxy luminosity.
These are both cD galaxies in nearby clusters which might have had rather
different mechanism of globular cluster formation, such as secondary formation
in the cooling flows
(Richer et al. 1993, but see also Holtzman et al. 1996 for an objection),
capture of globular clusters from other galaxies through tidal stripping
(C\^ot\'e, Marzke \& West 1998)
or the debris of cannibalized nucleated dwarf galaxies
(Bassino, Muzzio \& Rabolli 1994).
Apart from

\hbox{~}
\centerline{\psfig{file=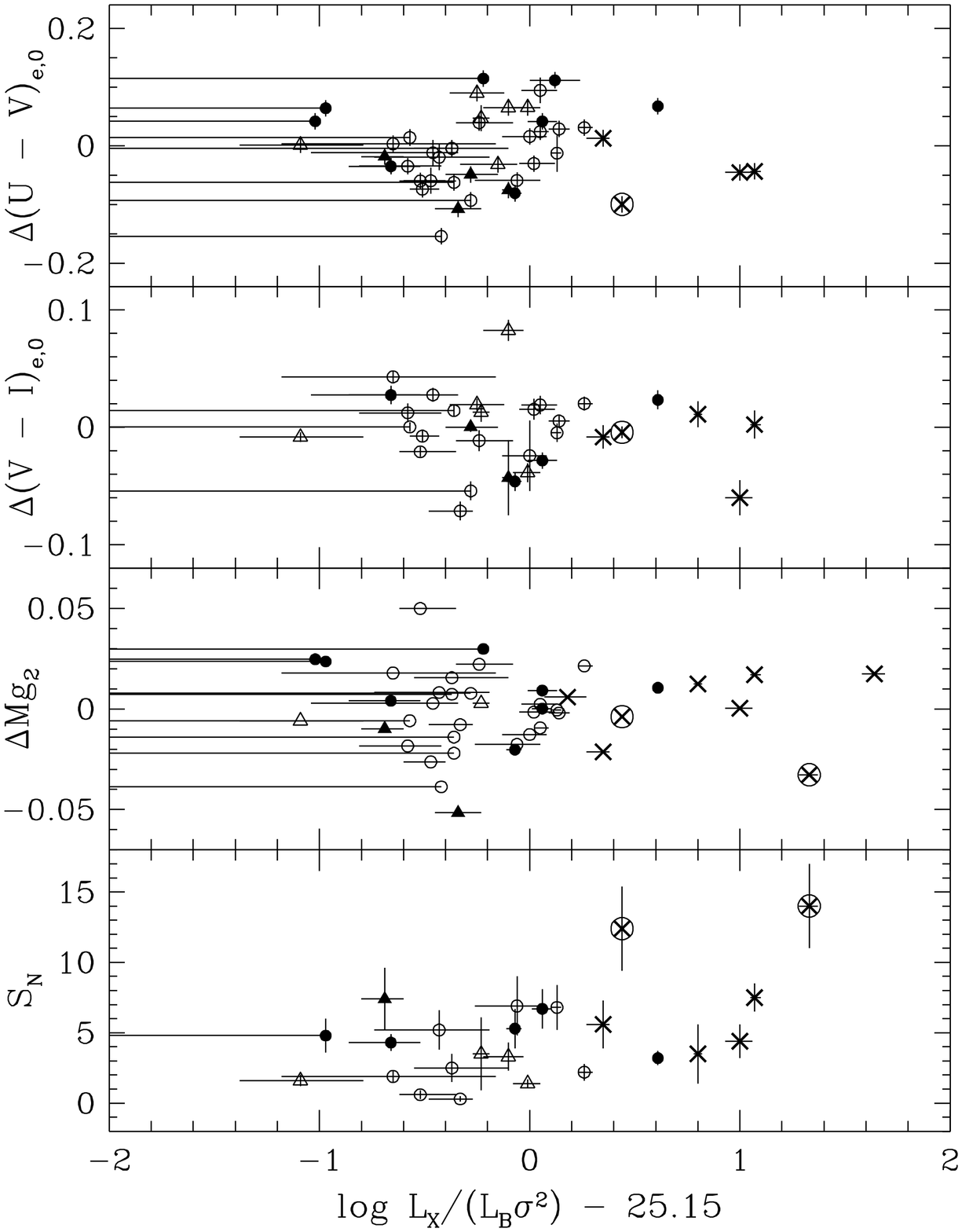,width=3.0in,angle=0}}

\vskip 2mm

\noindent{
\small \addtolength{\baselineskip}{-2pt} 
Fig.~4.\
Deviations from the $(U-V)$-$\sigma$ and $(V-I)$-$\sigma$ relations,
deviation from the
Mg$_2$-$\sigma$ relation, and $S_N$ (from top to bottom)
are plotted against the X-ray extent defined as
$E_X$$=$$\log L_X/(L_B\sigma^2)$$-$25.15.
Vertical error-bars (one sigma) are shown except for the Mg$_2$ index,
and the horizontal error-bars correspond to two sigma errors in $L_X$.
Crosses indicate the X-ray extended galaxies, and those surrounded by
big circles are the cD galaxies.
The galaxies in cluster environment are shown by filled symbols,
while those in the field or in small groups are shown by open symbols.
Ellipticals are indicated by circles, while S0's are indicated by triangles.

\vskip 5mm
\addtolength{\baselineskip}{2pt}
}

\noindent
these cD galaxies, the number of globular clusters seems
comparable between the X-ray extended galaxies and the X-ray compact ones.
The consequence of this result will be discussed in \S~4.

In order to compare the optical properties against the X-ray properties
in a more general manner, we use $E_X$ as a measure of the
X-ray extent or the depth of the potential (\S~2).
To subtract the underlying metallicity effect as a function of galaxy mass
(or $\sigma$) which is present in early-type galaxies
(eg., Kodama et al. 1998),
we measure the deviations from the color-$\sigma$ and the Mg$_2$-$\sigma$
relations (solid lines in Fig.~3) for each galaxy,
and plotted them against the X-ray extent ($E_X$) in Fig.~4.
The raw values of $S_N$ are re-plotted in the bottom panel .
There is no clear trend that any of the optical quantities scale
with the X-ray extent.
Our result is consistent with White \& Sarazin (1991) who found
no correlation between the excess of X-ray luminosity for a given optical
luminosity and the residual from the scaling relations in $U-V$ color
and Mg$_2$ index.

To test the above results statistically, we calculate the mean
deviations in four colors, the mean deviations in Mg$_2$ index,
and the mean $S_N$ for the X-ray extended galaxies and the X-ray compact ones
separately. The results are summarized in Table~1.
The standard deviation within each galaxy category is also given as a measure
of internal scatter. Since NGC~1399 and NGC~4486
have significantly larger $S_N$ values, we excluded these two cD galaxies
in calculating the average and the scatter of $S_N$ for
the X-ray extended galaxies.
We applied Welch's

\hbox{~}
\centerline{\psfig{file=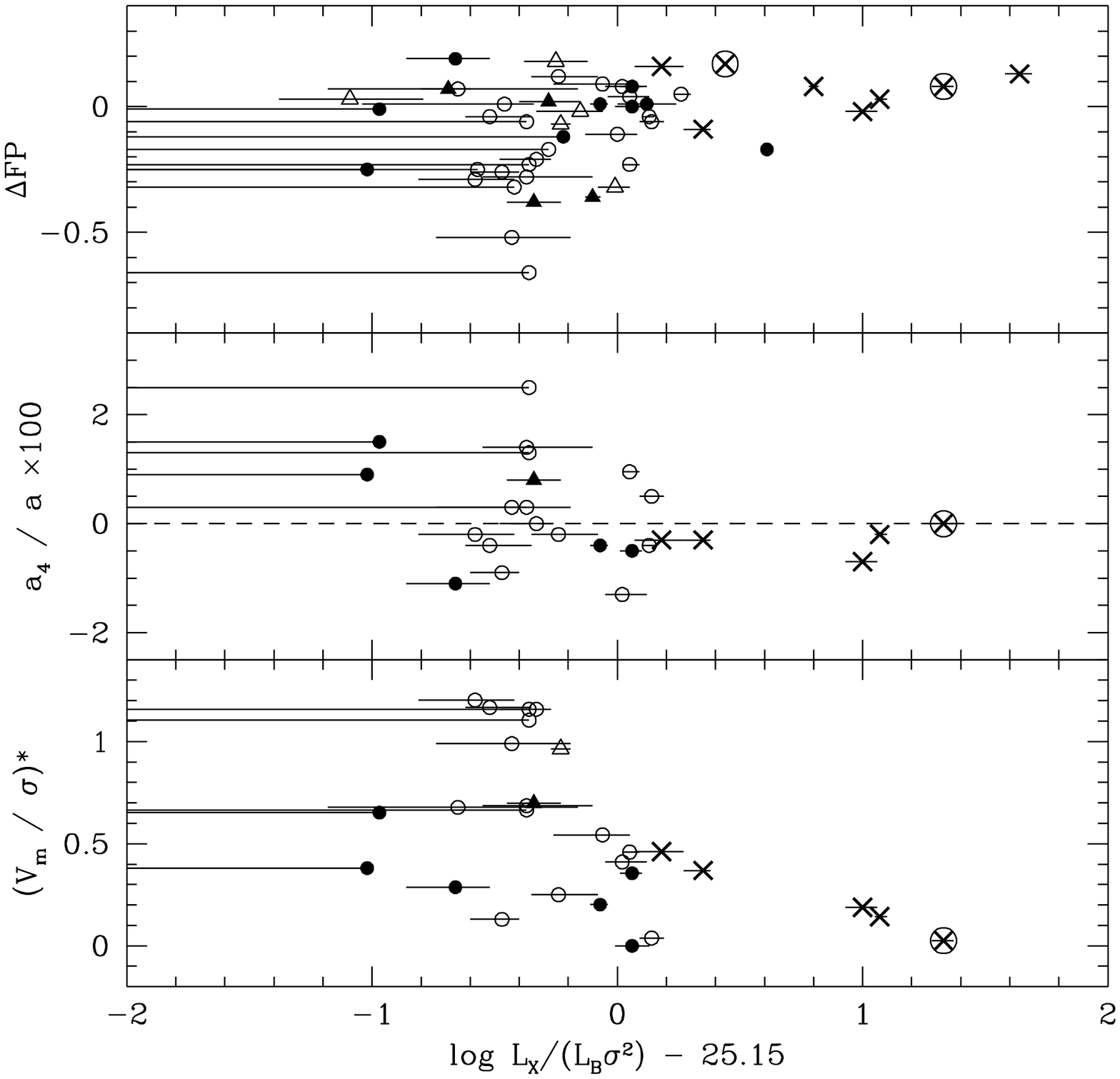,width=3.2in,angle=0}}

\vskip 2mm

\noindent{
\small \addtolength{\baselineskip}{-2pt} 
Fig.~5.\
Deviation from the Fundamental Plane ($\Delta$ FP), boxy/disky index
($a_4/a$), and velocity anisotropy index $(V_m/\sigma)^*$
(from top to bottom)
are plotted against the X-ray extent defined as
$E_X$$=$$\log L_X/(L_B\sigma^2)$$-$25.15.
Typical errors for ($a_4/a$$\times$100) and ($V_m/\sigma$)$^*$
are 0.25 and 10\%, respectively.
Holizontal error-bars correspond to two sigma errors in $L_X$.
Crosses indicate the X-ray extended galaxies, and those surrounded by
big circles are the cD galaxies.
The galaxies in cluster environment are shown by filled symbols,
while those in the field or in small groups are shown by open symbols.
Ellipticals are indicated by circles, while S0's are indicated by triangles.

\vskip 5mm
\addtolength{\baselineskip}{2pt}
}

\noindent
non-parametric statistical test and found that none of
these optical quantities show significant difference between the X-ray
extended galaxies and the compact ones. The probability of the difference
is always smaller than 90 per cent, and the hypothesis that we draw
both samples from the same parent group cannot be rejected.

If a difference of more than 0.078 in $U-V$ or 0.023 in Mg$_2$
would have been present between the two groups, it should have been detected
at the significance level in this statistical test.
This corresponds to the difference in stellar populations of only
$\Delta \log Z=0.1$ or $\Delta \log T=0.15$ for old galaxies
(Kodama \& Arimoto 1997).
Therefore the stellar populations of the early-type galaxies
should be homogeneous below this level despite the variety of X-ray extent.
The above upper limit for the metallicity difference is rather small compared
to the expected difference if the galaxies with the same stellar mass form
in various potential depth. We will discuss this point in \S~4.

We also plot the deviation from the fundamental plane, $\Delta{\rm FP}$,
in the top panel of Fig.~5 as a function of $E_X$.
$\Delta{\rm FP}$ is a measure of the difference of the dynamical mass-to-light
ratio from that of the `normal' early-type galaxies with the same dynamical
mass (Prugniel \& Simien 1996). This indicates the deviations in
stellar populations and/or dynamical mass including dark matter.
A positive value of $\Delta{\rm FP}$ corresponds to a larger mass-to-light
ratio. The X-ray extended galaxies generally have a high $\Delta{\rm FP}$,
while the X-ray compact ones show a considerable spread towards lower values.
Since we have found that the stellar populations are quite homogeneous
against $E_X$, the above result is indicative of the presence of more dark
matter in the X-ray extended galaxies, as expected.

Finally, we compare the isophotal shape and the velocity anisotropy against
the X-ray properties.
We first use the $a_4/a$ index, where $a_4$ is the fourth cosine coefficient of
the Fourier expansion of the deviations from a pure ellipse
and $a$ is the semi-major axis of the isophote (Bender \& M\"ollenhoff 1987).
A positive value of $a_4/a$ means a disky isophote,
while a negative value corresponds to a box-shaped isophote.
The other index, $V_m/\sigma$ is a ratio between maximum rotation velocity
($V_m$) and central velocity dispersion ($\sigma$), which is transformed
to the anisotropy index by taking into account the ellipticity ($\epsilon$) as:
\begin{equation}
(V_m/\sigma)^*=(V_m/\sigma)/ \sqrt{\epsilon/(1-\epsilon)},
\end{equation}
following Bender (1988).
There are clear trends that the X-ray extended galaxies have a negative 
$a_4/a$ and small $(V_m/\sigma)^*$ with relatively small scatters.
These effects should partly come from the dependence of $a_4/a$ and
$(V_m/\sigma)^*$ on galaxy luminosity (Bender 1988, Bender et al. 1989),
because the X-ray extended galaxies are generally bright. 
However, it is notable that all of the X-ray extended galaxies have 
boxy shapes and weak rotations.
These findings are similar to what Bender (1988) and Bender et al.
(1989) found; ie., 
$a_4/a$ index correlates with the X-ray luminosity excess that comes from the
surrounding hot gas halos and also with the velocity anisotropy.
Some attempts have been made to understand this kinematical dichotomy of
elliptical galaxies in the context of galaxy-galaxy merging using dynamical
simulations (Bekki \& Shioya 1997;
Bruckert et al. 1999). Bruckert et al. (1999) showed that major
mergers produced boxy ellipticals with anisotropic velocity, while minor
mergers produced the disky ones.
Considering together that the X-ray extended galaxies are located at the local
density peaks, they are possibly the products of major mergers.

After all, although the clear dichotomy in X-ray properties correlates with
the isophotal shape and the velocity structure of
galaxies, the stellar populations and globular cluster properties are
still found to be quite homogeneous.

\begin{deluxetable}{cccc}
\tablewidth{6.5in}
\small
\tablenum{1}
\label{table-1}
\tablecaption{Difference of the indices between the X-ray extended galaxies
and the X-ray compact ones.}
\tablecomments{
From top to bottom, the average values are shown for
the deviations from the scaling relations of the four colors
and Mg$_2$ against $\log\sigma$, $S_N$, deviation from the
fundamental plane, boxy/disky index, and anisotropy index.
Standard deviations are given following the $\pm$ signs as the measure of
scatters around the average values.
The two cD galaxies, NGC~1399 and NGC~4486, are excluded in calculating
the averaged $S_N$ value and the scatter for the X-ray extended galaxies.
}
\tablehead{
\colhead{Index} & \colhead{X-ray} & \colhead{X-ray} & \colhead{probability of}\\
\colhead{} & \colhead{compact} & \colhead{extended}  & \colhead{the difference}}
\startdata
$\Delta(U-V)_{e,0}$   & $0.00 \pm 0.07$ & $-0.04 \pm 0.05$ & $<$90~\% \\
$\Delta(B-V)_{e,0}$   & $0.00 \pm 0.03$ & $-0.01 \pm 0.02$ & $<$80~\% \\
$\Delta(V-R)_{e,0}$   & $0.00 \pm 0.01$ & $-0.01 \pm 0.02$ & $<$80~\% \\
$\Delta(V-I)_{e,0}$   & $0.00 \pm 0.03$ & $-0.01 \pm 0.03$ & $<$50~\% \\
$\Delta$ Mg$_2$       & $0.00 \pm 0.02$ &  $0.00 \pm 0.02$ & $<$50~\% \\
$S_N$ & $3.80 \pm 2.33$ & $5.25 \pm 1.73$ & $<$80~\% \\
\hline
$\Delta$ FP & $-0.10 \pm 0.19$ & $0.07 \pm 0.09$ & $>$99~\% \\
$a_4/a$$\times$100 & $0.25 \pm 0.98$ & $-0.30 \pm 0.26$ & $>$95~\% \\
$(V_m/\sigma)^*$ & $0.59 \pm 0.40$ & $0.24 \pm 0.18$ & $>$99~\%
\enddata
\end{deluxetable}
\vskip -10mm

\section{Discussion}

We estimate how much metallicity difference is expected
between the X-ray extended galaxies and the X-ray compact ones
at a given stellar mass (or $\sigma$) if the dichotomy of the potential
structure has already been established at the time of star formation.
The hydrostatic equilibrium for the ISM gives the galaxy potential of
\begin{equation}
\Phi \propto T\left(\frac{\log \rho}{\log r}+\frac{\log T}{\log r}\right),
\end{equation}
where $\rho$ is the density of matters including the dark matter.
Since the density gradient $\log \rho/\log r$ should not differ
much between the two galaxy categories (X-ray extended and the X-ray
compact ones) within optical radius, and the ISM temperature gradient
$\log T/\log r$ is negligible compared to the density gradient
(Forman, Jones \& Tucker 1985; Trinchieri et al. 1994;
Boute \& Canizares 1994),
the potential $\phi$ approximately scales with the ISM temperature $T$.
Given that there is a factor of 2 difference in $T$ within 4~$r_e$ between
the two galaxy categories for a given $\sigma$ (\S~2),
the X-ray extended galaxies could have experienced twice as many supernova
explosions and recycled twice as much metals into the same amount of stars
before the gas is expelled.
Therefore we could expect that the mean stellar metallicity of the X-ray
extended galaxies is twice as high as that of the X-ray compact ones
for a given $\sigma$.
This big difference in metallicity should have been detected in the
statistical test in \S~3, if present. As opposed to what we expect, however,
we do not detect any significant difference in stellar populations
between the two.
This means that the potential structure of early-type
galaxies during the major star formation epoch was quite different from
what it is today.
Gravitational potential must have been homogeneous.
It must not have produced more than only 0.1~dex difference
in mean stellar metallicity at a given stellar mass.
Later on, after the epoch of major star formation,
some galaxies became incorporated into larger scale potentials by
infalling into the bottom of the local potential and/or by accumulating
the surrounding materials and augmenting their gravitational potential,
which eventually resulted in the variety of X-ray extent seen at the
present-day.

This picture is consistent with what people have found as to the
formation of early-type galaxies:
Most of the `stars' in early-type galaxies should form at significantly high
redshifts ($z>2$) (eg., Bower, Lucey \& Ellis 1992; Ellis et al. 1997;
Stanford, Eisenhardt \& Dickinson 1998; Kodama et al. 1998;
van Dokkum et al. 1998; Kodama, Bower \& Bell 1999),
while the `mass' of early-type galaxies can successively grow due to
the accretion and/or merging even well below $z<2$ in the course of
hierarchical assembly (eg., Kauffman 1996; Baugh et al. 1998; Bower, Kodama
\& Terlevich 1998; van Dokkum et al. 1999).

From our results, we speculate that the chemical abundance of ISM in
$\alpha$-elements (such as O, Mg, Si, and S which come mainly from
SN Type II) should be quite uniform at a fixed stellar mass
of the central galaxy regardless of its X-ray extent.
The individual potential structure of a galaxy is independent of the larger
scale potential and should be similar at the time of major star formation.
Davis \& White (1996) and Loewenstein et al. (1994) claimed
that galaxies with lower ISM temperature or compact X-ray halos
tend to have lower chemical abundance of the ISM.
However, it is not yet clear whether there is such a correlation for
the $\alpha$-elements abundance (Matsushita, Makishima \& Ohashi 2000).
The new X-ray satellite XMM will solve this problem.
On the other hand, the contribution from SN Type Ia can be much different
because the potential structure would vary at the time of a SN Ia
explosion due to the time delay of the explosion (Yoshii, Tsujimoto \&
Nomoto 1996).
The X-ray extended galaxies would have acquired deeper potential wells by then,
keep the SNIa ejecta more efficiently and hence show relatively iron
enhanced ISM abundance compared to the X-ray compact ones.
This is what we actually observe by ASCA.
Although the ISM abundance of the X-ray compact galaxies is still
uncertain, if we assume the same $\alpha$-element abundance as
the X-ray extended systems, the Fe abundance is about a factor of 2 smaller
than that of the X-ray extended objects
(Matsushita, Makishima \& Ohashi 2000).

Considering that the X-ray extended galaxies are sitting in the center of
larger scale potential structures and that dynamical friction drives satellite
galaxies towards the center,
these galaxies are more likely to be produced by galaxy mergers.
The fact that the X-ray extended galaxies tend to have boxy shapes and
weak rotation might support this hypothesis.
If this is really the case, and if a considerable number of new globular
clusters form during galaxy mergers as
suggested by Zepf \& Ashman (1993), we should expect higher $S_N$ for the
X-ray extended galaxies on average.
However, there is no significant difference in $S_N$ except for
the cD galaxies. This may imply that, apart from the cD galaxies,
the secondary globular clusters do not generally form by recent
galaxy mergers,
and that most of the globular clusters around the X-ray extended galaxies
are likely to form very early when the major star formation takes place
in their host galaxies.

\section{Conclusions}

The stellar population makeup in early-type galaxies does not correlate with
their present-day global potential structure.
Early-type galaxies form stars at early epoch in their own potential wells
independently, and some of the galaxies become incorporated into larger
scale potential structures (clusters/groups) later on. This idea naturally
explains the homogeneity of the stellar populations despite the variety of
X-ray properties.

\acknowledgments

This work was supported by
the Japan Society for the Promotion of Science (JSPS) through
its Research Fellowships for Young Scientists.
We thank K. Pimbblet for carefully reading the paper and polishing up
English as well as giving us some useful comments.

\end{document}